\documentclass[traditabstract, bibyear]{aa} 
\usepackage{graphicx}
\usepackage{txfonts}
\usepackage{color}
\usepackage{amstext}
\usepackage{answers}

\begin{document}

   \title{The secondary transit of the hot Jupiter WASP-121b at 2~$\mu$m}

   \author{G\'eza Kov\'acs\inst{1}
          \and 
	  Tam\'as Kov\'acs\inst{2}}

   \institute{Konkoly Observatory of the Hungarian Academy of Sciences, 
              Budapest, 1121 Konkoly Thege ut. 13-15, Hungary \\
              \email{kovacs@konkoly.hu}
	  \and
              Institute of Theoretical Physics, E\"otv\"os University, 
              Budapest, 1117 P\'azm\'any P\'eter s\'et\'any 1A
             }

   \date{Received September ?? 2018 / Accepted ?? 2018}

 
  \abstract
{Ground-based observations of the secondary transit in the 2MASS K band 
are presented for the hot Jupiter WASP-121b. These are the first 
occultation observations of an extrasolar planet carried out with 
an instrument attached to a $1$~m-class telescope (SMARTS' $1.3$~m). 
We find a highly significant transit depth of $(0.228\pm 0.023)$\%. 
Together with the Hubble Space Telescope near infrared emission 
spectrum, current data support more involved atmosphere models with 
species producing emission/absorption features, rather than simple 
smooth black body emission. Analysis of the time difference between 
the primary and secondary transits and the durations of these events 
yield an eccentricity of $e=0.0207\pm 0.0153$, which is consistent 
with the earlier estimates of low/zero eccentricity, but with a smaller 
error. Together with the existing K-band data on other systems, 
WASP-121b lends further support to the lack of efficient heat transport 
between the day and night sides for nearly all Hot Jupiters.}

   
\titlerunning{The secondary transit of WASP-121b}
\authorrunning{Kovacs, G. \& Kovacs, T.}
   \maketitle
%

%
%
\section{Introduction}
When combined with other pieces of information (such as planet mass), 
the low thermal radiation of extrasolar planets is a direct evidence 
of their substellar nature. In addition to this independent verification, 
measuring the radiation spectrum yields a wealth of information on atmospheric 
structure and basic orbital parameters. Due to their low temperatures (relative 
to their host stars), the best chance of detection is obviously in the infrared. 
From the first detection, employing the mid-infrared instrument of the Spitzer 
Space Telescope by Deming et al.~(\cite{deming2005}), many systems have been 
observed not only by space-based, but also by ground-based instruments attached 
to 4-meter class telescopes. Here we report the multiple detection of the 
secondary transit\footnote{Throughout this papers, we use also the word   
`occultation' for the event of secondary transit.} of the Hot Jupiter (HJ) 
WASP-121b in the 2MASS K band by A Novel Dual Imaging CAMera (ANDICAM) attached 
to the $1.3$~m telescope of the SMARTS Consortium\footnote{For additional 
details on the instrument, access images, raw and systematics corrected light 
curves, please visit \\  
\url{http://www.astro.yale.edu/smarts/1.3m.html} \\ 
\url{http://archive.noao.edu/search/query/} \\ 
\url{http://cdsweb.u-strasbg.fr/} .}.  

The transiting extrasolar planetary system WASP-121 was discovered by 
Delrez et al.~(\cite{delrez2016}) using the wide-field telescopes of the SuperWASP 
project (Pollacco et al.~\cite{pollacco2006}, see also Anderson et al.~\cite{anderson2018} 
for the latest update). The analysis of these and the subsequent spectroscopic 
followup observations revealed that WASP-121b is a very hot Jupiter, with a 
maximum photospheric temperature above $3000$~K. This is not surprising, 
since the planet orbits an F star rather close, with a period of $1.27$~days. 
The close orbit, the extended planet's radius\footnote{The radius quoted here is 
the one that appears in the Abstract of the discovery paper. We use the radius 
based on the HST measurements of Evans et al.~(\cite{evans2017}) -- see Sect.~4.} 
of $R_{\rm p}=1.87~R_{\rm J}$ with standard mass of $M_{\rm p}=1.18~M_{\rm J}$ 
imply rather strong tidal dissipation, leading to Roche-lobe filling and then to 
a speedy disruption within some hundred million years, assuming a stellar tidal 
dissipation factor $Q_{\rm star} < 10^8$ (see Delrez et al.~\cite{delrez2016}). 
In addition, the planet has probably experienced a strong dynamical interaction 
with some nearby third body, as can be inferred from the large projected spin-orbit 
angle of $258^{\circ}\pm5^{\circ}$ (see Delrez et al.~\cite{delrez2016}). 
Interestingly, secondary transits were observed multiple times by the 
same authors in the Sloan-z' band by the $60$~cm TRAPPIST telescope -- 
to our knowledge, this is the first occultation detection from the ground 
by a telescope of this size. The significant detection of the occultation 
depth of a mere $(0.060\pm0.013)$\% resulted in the first direct estimation 
of the planet temperature. Important followup observations (both during 
the primary and secondary transits) have been made by 
Evans et al.~(\cite{evans2016, evans2017, evans2018}) by using the Hubble 
Space Telescope's Wide Field Camera 3 in the near infrared, Spitzer's IRAC 
detector at $3.6$~$\mu$m and HST/STIS in the UV. These data indicate a weak 
$H_2O$ emission during occultation and absorption during transit, 
implying temperature inversion due to some high-altitude absorber. In spite 
of the successful fit of the HST emission spectrum, and quite currently the 
transmission spectrum observed by the same instrument, the authors caution 
for the non-uniqueness of the solution (e.g., type of absorber, precision 
of the fit at different wavelengths, etc.). These issues are not unique to  
WASP-121b, they are also present in other very hot Jupiters (e.g., WASP-33b, 
Kepler-13Ab -- see Parmentier et al.~\cite{parmentier2018}). In spite of 
the considerable progress made in the past ten years, there is a substantial 
lack of understanding the relations between the physical parameters of the 
systems and the thermal properties of their planets (see the uniform analysis 
by Adams \& Laughlin~\cite{adams2018} of 10 systems with full infrared phase 
curves).      

The purpose of this work is to add a flux value to the emission spectrum of 
WASP-121b at a single waveband and thereby increase the number of constraints 
on future atmosphere modeling of the planet. Furthermore, timing estimates are 
presented to give more stringent limits on the orbital eccentricity, a 
valuable parameter for the analysis of the dynamical history of the system. 

%
%
\section{Observations and the method of analysis}
Photometric observations in the 2MASS K and Cousins I bands (effective 
wavelengths of $2.2$~$\mu$m and $0.8$~$\mu$m, respectively) have been made 
by using the ANDICAM instrument in a beam-splitting mode, allowing simultaneous 
data acquisition in the two bands.\footnote{Unfortunately, the signal -- hampered 
by weather and instrumental limitations -- in the Cousins I band was too weak 
to yield any useful planet atmospheric constraint, so we decided not to deal 
with it.} On each night, the target was monitored continuously, by allowing 
ample amount of pre- and after-event time (permitted by the actual sky position) 
to reliably fix the out-of-transit (OOT) baseline for the event lasting almost 
for $3$~hours. An exposure time of $15$--$20$~s was used, resulting 
in an overall cadence of $70$--$500$~s, due to the overheads, related to 
read-outs, varying movements due to dithering and other data acquisition steps. 
The observing log with some associated parameters is given in Table~\ref{obs_log}. 

%
%
\begin{table}[!h]
  \caption{List of secondary transit observations of WASP-121b in the near infrared.}
  \label{obs_log}
  \scalebox{1.0}{
  \begin{tabular}{cccc}
  \hline
  Date [UT] & $T_{\rm occ}$ [HJD] & $T_{\rm tot}$ [hours] & $N_{\rm tot}$ \\ 
 \hline
 02/26/2016 & 2457444.64855   &   4.86   & 177 \\
 01/11/2017 & 2457764.65485   &   6.13   & 203 \\
 01/25/2017 & 2457778.67903   &   6.64   & 205 \\ 
\hline
\end{tabular}}
\begin{flushleft}
{\bf Notes:}\\
\vbox{\small 
$T_{\rm occ}$ stands for the time of the center of occultation as estimated 
from the ephemeris given by Delrez et al.~(\cite{delrez2016}) for the primary 
transit in their Table~4. $T_{\rm tot}$ and $N_{\rm tot}$ are the total 
observing time and data points gathered.} 
\end{flushleft}
\end{table}
For the K-band observations dithering was used to decrease the higher 
sensitivity against detector non-uniformity in the near infrared. We 
found this method useful, as we did not have a priori information on 
pixel sensitivity. Admittedly, this method has also some risk, since 
by testing different parts of the CCD, we may bump into bad positions, 
leading to light curves of larger scatter associated with the particular 
dither position. All in all, we think that our strategy has proven to 
be useful and led to a higher quality result at the end. 

\begin{figure}
 \vspace{0pt}
 \centering
 \includegraphics[angle=-0,width=75mm]{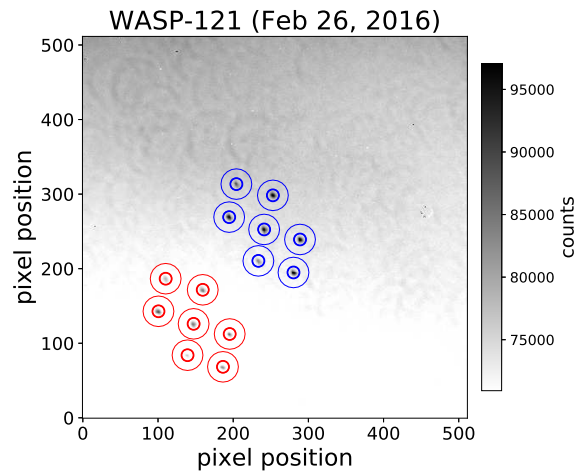}
 \caption{Dithering pattern used during the near infrared observations. 
 The image shows the $2.4'\times2.4'$ FOV of the ANDICAM near infrared 
 camera, attached to the $1.3$~m telescope of the SMARTS Consortium. 
 The target (WASP-121=2MASS 07102406-3905506) is in the middle, the 
 comparison star 2MASS 07102364-3905561 is in the lower left corner. 
 North is on the top West is on the left. Circles around the target 
 and the comparison star show the aperture sizes used to estimate the 
 star and background fluxes.}
\label{dither}
\end{figure}

By stacking several images, we show the dither pattern in Fig.~\ref{dither} 
for one of the nights. The number of dither positions changed from night to 
night, and their durations were also not the same. The image (that has already 
been corrected for flat field) spectacularly exhibits sequences of rings, 
reminiscent of the trace of some earlier drops of dew. In spite of their high 
visibility, their effect has been proven to be less damaging for the data 
quality than the varying pixel sensitivity (that is considerable more 
difficult to spot, because they lack the type of spatial correlation 
the rings have).  

%
%

In producing the photometric time series to be used in the derivation of the 
basic occultation parameters, we proceed as follows. First we compute simple 
relative fluxes at various, but fixed circular apertures from $10$ to $20$ pixel 
radii with an increment of $2$ pixels. After a lot of experimenting, and 
inspecting the final product of the full detrending procedure to be described 
below, we find that the aperture with the pixel radius of $16$ yields a light 
curve with the smallest scatter. All results presented in this paper refer to 
the above aperture size. 

%
%
\begin{figure}
 \vspace{0pt}
 \centering
 \includegraphics[angle=-90,width=75mm]{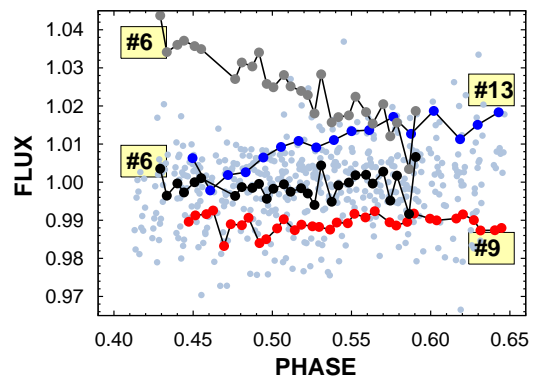}
 \caption{Simple photometric flux ratios ordered by the orbital period 
          (pale dots). Some dithers are annotated to see the nightly 
	  trends (or the lack of them, i.e., \#9). Dither \#6 (gray dots)  
	  is plotted also after employing zero point shift and detrending 
	  by the position vector (black dots, see Sect.~2 for details).}
\label{raw_lc}
\end{figure}

By using the relative fluxes (target over comparison star flux, hereafter raw flux) 
and folding the data with the orbital period, we can examine if we can see 
some sign of an occultation event. The result is shown in Fig.~\ref{raw_lc}. 
The pale dots show that the raw fluxes are very noisy, and the event with the 
expected depth of $0.1$--$0.2$\% is hopelessly buried in the noise. We can 
find out the reason of this somewhat unexpected high level of noise by 
examining the individual light curves (LCs) associated with the various 
dither positions\footnote{Please note that all dither positions are counted 
and their indices increase toward more recent nights of observation.}. 
Indeed, we see from the highlighted LCs that there is a strong dependence 
on the dither position, leading to both zero point shifts and nightly trends. 
Therefore, (not entirely unexpectedly), we must employ some detrending 
method that is likely the cause of the trends and zero point shifts. The  
detrending step is vital and therefore quite common in the extraction of 
planetary signals in general, and, in particular, in the derivation of 
wavelength-dependent transit depths for the exquisite accuracy needed to 
estimate emission or transmission spectra (e.g., Stevenson et al.~\cite{stevenson2012};  
Kreidberg et al.~\cite{kreidberg2015}).    

It is well-known that ground-based instruments detect stellar light 
deformed by the multiplicative noise and systematics originating from 
the Earth's atmosphere and from the environment/instrument. In addition, 
we also have an additive noise source, coming from the sky background 
%
\begin{eqnarray}
\label{eq:1}
F & = & F_0\times T_{\rm atm}\times T_{\rm env} + F_{\rm bg} \hskip 1mm ,
\end{eqnarray}     
where $F$ is the detected, $F_0$ is the true stellar flux. The transmission 
functions of the Earth's atmosphere and the instrument are denoted by 
$T_{\rm atm}$ and $T_{\rm env}$, respectively, whereas the background flux 
by $F_{\rm bg}$. In traditional photometric reductions the atmospheric 
and instrumental effects are filtered out with the aid of comparison stars 
near the target, by using the assumption of the close similarity in the 
transmission functions for the target and its neighboring companions. 
However, when higher accuracy is required, this method usually fails, because 
of the lack of complete equivalence between the transmission functions of 
the target and the comparison stars (and, for faint targets, there is also 
the issue of the presence of the additive background noise). 

Due to the lack of obvious exact solution of the problem (similarly to the 
methodology followed in other studies, e.g., Bakos et al.~\cite{bakos2010}, 
Delrez et al.~\cite{delrez2016}), we opt to an approximate one. Here we take 
the logarithm of the target to comparison star flux ratios $F/F^{\rm c}$, 
and fit the data with the linear combination of the presumed signal and 
certain external photometric parameters (e.g., position, PSF width). 
In addition, we treat each LC of the different dither positions individually, 
with particular zero points and trends (but with the same underlying signal). 
That is, we Least Square minimize the following expression 
%
\begin{eqnarray}
\label{eq:2-3}
\mathcal D & = & \sum_{j=1}^M  \sum_{i=1}^{N_{\rm j}} w_{\rm j}\bigg[\log\Bigg({F_{\rm j}(i) \over F_{\rm j}^{\rm c}(i)}\Bigg) - E_{\rm j}(i)\bigg]^2 \hskip 2mm ,\\
E_{\rm j}(i) & = & a_{0,\rm j} + a_{x,\rm j}X_{\rm j}(i) + a_{y,\rm j}Y_{\rm j}(i) + A \log(F_{\rm trap}(i)) 
\hskip 1mm .
\end{eqnarray}   
Here all data are sorted by the orbital phase. We assume that there are $M$ dither 
positions altogether with $N_{\rm j}$ data points at the $j$-th dither. Since our 
extensive tests showed that neither arbitrary polynomial nor additional external 
parameters are needed to reach a high signal-to-noise ratio ($S/N$) detection, we 
use only the pixel position components $(X,Y)$ of the target to correct for 
instrumental effects. The stellar flux during the occultation is approximated by 
a trapezoidal function $F_{\rm trap}$ with fixed ingress/egress time, duration and 
transit center of $0.015$, $0.120$ and $2457764.65485$ days, respectively, 
corresponding to those given by Delrez et al.~(\cite{delrez2016}). The weights $\{w\}$ 
are constant for the same dither index and proportional to the reciprocal of the 
variance of the residuals around the best-fitting trapezoidal. Since the solution 
is not known, the weights are iterated during the process of solution. At the end, 
the data are converted back to relative intensities, with an OOT normalization of 
$1.0$ for the fitted trapezoidal. The error of the occultation depth is computed 
as follows 
%
\begin{eqnarray}
\label{eq:4}
\sigma(\delta_{\rm occ}) = \Bigg(1 - {N_{\rm p} \over N}\Bigg)^{-1/2}\Bigg({s_{14} \over N_{14}^2} + {s_{\rm oot} \over N_{\rm oot}^2}\Bigg)^{1/2}
\hskip 1mm ,
\end{eqnarray}   
where $s_{14}$ and $s_{\rm oot}$ are sums of the squared residuals in the in- and 
out-of-transit phases, respectively. Akin to these are the number of data points 
$N_{14}$ and $N_{\rm oot}$. The factor in front (with $N_{\rm p}$ number of the 
parameters fitted to $N$ total number of data points) is for the debiasing of 
the error due to the decrease of the degrees of freedom, because of parameter 
fitting. The $S/N$ of the detection is the ratio of the average transit depth to 
the above error 
%
\begin{eqnarray}
\label{eq:5}
S/N = {1 \over \sigma(\delta_{\rm occ})} {1 \over N_{14}}\sum_{i=1}^{N_{14}} F_{\rm trap}(i) 
\hskip 1mm .
\end{eqnarray}   
%

%
%
\section{Occultation parameters}
\begin{figure}
 \vspace{0pt}
 \centering
 \includegraphics[angle=-90,width=85mm]{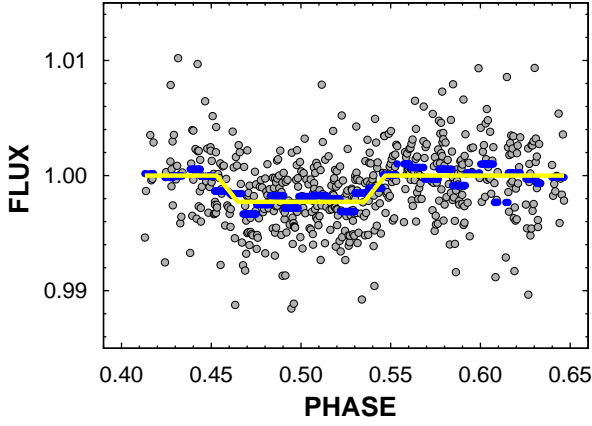}
 \caption{Systematics-filtered folded flux ratios normalized to $1.0$ 
          in the out-of-transit part. Average fluxes (in $30$ phase bins) 
	  are shown by blue dashes, the best fitting trapezoidal secondary 
	  transit approximation is plotted by yellow continuous line.}
\label{final_lc}
\end{figure}

First we fix all secondary transit parameters (but the occultation depth) 
by assuming circular orbit and the validity of the parameters derived 
for the primary transit by Delrez et al.~(\cite{delrez2016}). Following 
the procedure described in Sect.~2, we compute the best fitting occultation 
depth under various conditions, concerning the number of points clipped 
and the dither LCs omitted. The result is shown in Table.~\ref{occ_depth}. 
Except perhaps for the extreme choices of data trimming parameters 
($N_{\rm dith}^{\rm cut}$ and $N_{\sigma}^{\rm cut}$), the occultation 
depth is relative stable. To avoid too sparsely populated dither LCs, 
and not to `overtrim' the data, we opt for the case of $N_{\rm dith}^{\rm cut}=10$ 
and $N_{\sigma}^{\rm cut}=4$. The folded LC obtained in this way is shown 
in Fig.~\ref{final_lc}. The resulting secondary transit depth is $0.00228\pm 0.00023$. 

%
%
\begin{table}[!h]
  \caption{Occultation depths for WASP-121 in the 2MASS K band}
  \label{occ_depth}
  \scalebox{1.0}{
  \begin{tabular}{ccccccc}
  \hline
  $N_{\rm dith}^{\rm cut}$ & $N_{\sigma}^{\rm cut}$ & $N_{\rm par}$ & $N$ & $S/N$ & $\sigma_{\rm fit}$ & $\delta_{\rm occ}$ \\ 
 \hline
 \phantom{0}0 & $\infty$ & $70+\phantom{0}0$  & $585$ & $8.1$ & $0.00272$ & $0.00212$\\
 \phantom{0}0 & $5$      & $70+\phantom{0}8$  & $585$ & $8.4$ & $0.00264$ & $0.00215$\\
 \phantom{0}0 & $4$      & $70+11$ & $585$ & $8.8$ & $0.00264$ & $0.00224$\\
 \phantom{0}0 & $3$      & $70+25$ & $585$ & $9.2$ & $0.00256$ & $0.00227$\\
 10 & $\infty$ & $61+\phantom{0}0$  & $561$ & $8.2$ & $0.00275$ & $0.00217$\\
 10 & $5$      & $61+\phantom{0}8$  & $561$ & $8.5$ & $0.00266$ & $0.00220$\\
 10 & $4$      & $61+11$ & $561$ & $8.9$ & $0.00266$ & $0.00228$\\
 10 & $3$      & $61+25$ & $561$ & $9.3$ & $0.00257$ & $0.00231$\\
 15 & $\infty$ & $52+\phantom{0}0$  & $520$ & $8.2$ & $0.00281$ & $0.00231$\\
 15 & $5$      & $52+\phantom{0}8$  & $520$ & $8.6$ & $0.00270$ & $0.00233$\\
 15 & $4$      & $52+11$ & $520$ & $8.9$ & $0.00270$ & $0.00243$\\
 15 & $3$      & $52+25$ & $520$ & $9.3$ & $0.00260$ & $0.00244$\\
\hline
\end{tabular}}
\begin{flushleft}
{\bf Notes:}\\
\vbox{\small 
$N_{\rm dith}^{\rm cut}$ is the lower limit on the number of data points 
per dither position. $N_{\sigma}^{\rm cut}$ is the number of standard 
deviations used in the clipping of the data points. $N_{\rm par}$ is the 
number of parameters fitted, plus the number of data points omitted 
($N_{\rm p}$ in Eq.~\ref{eq:4} includes both of these). Items in the 
last two columns (unbiased estimates of the standard deviation of the 
residuals and occultation depth) refer to the OOT$=1$ normalization 
as described in Sect.~2.} 
\end{flushleft}
\end{table}
%

%
%

\begin{figure}
 \vspace{0pt}
 \centering
 \includegraphics[angle=-90,width=85mm]{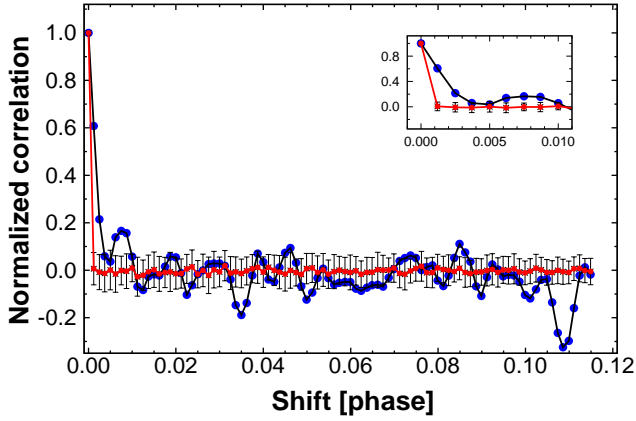}
 \caption{Blue dots: autocorrelation function (ACF) of the residuals of the 
          trapezoidal fit to the final dataset shown in Fig.~\ref{final_lc}. 
	  Red dots: ACF of generated uncorrelated noise. Error bars are for 
	  the standard deviations of the the ACF values of the random datasets. 
	  The time lag is given in the units of the orbital period. The inset 
	  shows the immediate neighborhood of ACF at zero time shift (given 
	  in the units of the orbital period).}
\label{acf}
\end{figure}

To check the level of the systematics filtering, we compute the 
autocorrelation function (ACF) of the residuals after subtracting 
the best-fitting trapezoidal as shown in Fig.~\ref{final_lc}. In the 
units of the orbital period, the ACF is computed with steps of 
$0.00123$ up to $0.115$, i.e. close to the length of the full  
transit event. As a sanity check, we also compute the ACF for many 
Gaussian white noise realizations. The result is shown in Fig.~\ref{acf}. 
We see that the residuals are almost uncorrelated. The basic 
correlation length is under $\sim 0.005$ in the units of the orbital 
period. This value is less than one half of the ingress duration. It 
seems that the de-correlation method applied yields nearly white noise 
residuals, supporting the validity of the pure statistical error estimation 
given by Eq.~\ref{eq:4}. We also note that similar short-time-scale 
correlations are observable in other studies dealing with systematics, 
in particular in the analysis of the HST data by Evans et al.~(\cite{evans2017}). 

%
%

Although the noise is rather high, the relative large number of data points led 
to a high-$S/N$ detection. Therefore, it is tempting to see if our assumption on  
the applicability of the primary transit parameters is really held, and if there 
is a chance to further constrain the eccentricity by the best-fitting occultation 
center and event duration. To this aim we map the quality of the fit as a function 
of $\Delta T_{\rm c}$ (tested occultation center time minus the one calculated 
from the primary transit with the assumption of circular orbit) and $t14$ 
(occultation duration). 

In addition to our data, to examine further the issue of eccentricity, 
the secondary transit data of Delrez et al.~(\cite{delrez2016}) are also 
investigated. Since the observations were made in the Sloan z' band, the 
signal is considerably shallower than in the 2MASS K ($K_{\rm s}$) band. 
Nevertheless, the number of data points ($6260$ flux measurements on seven 
nights) compensates for this, and yields a confident detection of $S/N=7.5$, 
with $\delta_{\rm occ}=0.000697\pm0.000081$ and a residual standard 
deviation of $0.003190$. This depth is larger by $0.000096$ than the one 
derived by Delrez et al.~(\cite{delrez2016}), but the difference is within 
$1\sigma$, and could be accounted for by the lower number of detrending 
parameters used in our code. We found it satisfactory to use only the pixel 
coordinates, and avoid to correct with a polynomial and other parameters, 
since these do not yield an appreciable improvement in the quality of the 
fit, and, in addition, may lead to a depression of the occultation depth. 

%
%
\begin{figure}
 \vspace{0pt}
 \centering
 \includegraphics[angle=-90,width=85mm]{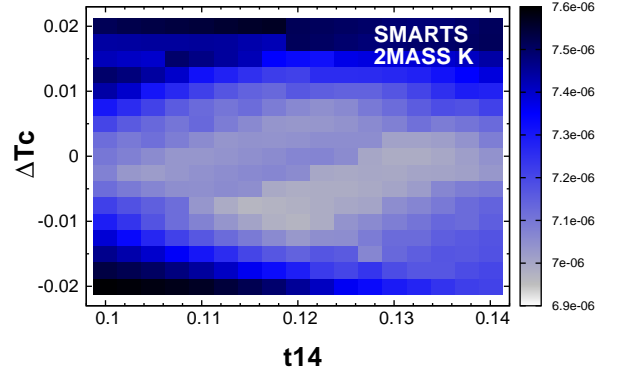}
 \caption{Intensity plot for the unbiased estimate of the variance of the residuals 
        between the data and the occultation model scanned in the parameter space 
	of the displacement of the occultation center $\Delta T_{\rm c}$ and the 
	duration of the event $t14$. We employ iterative $4\sigma$ clipping to find 
	the best solution for each parameter combination.} 
\label{t14-tc_smarts}
\end{figure}
%

%
%
\begin{figure}
 \vspace{0pt}
 \centering
 \includegraphics[angle=-90,width=85mm]{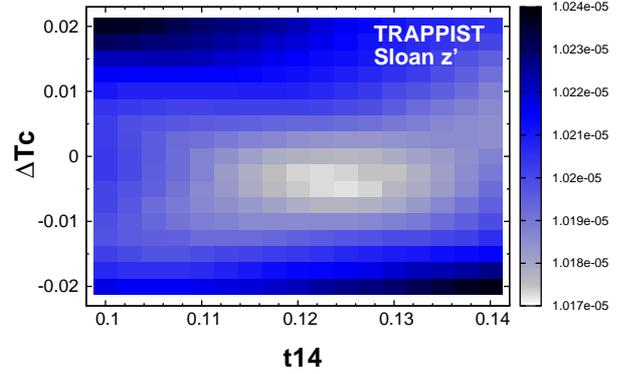}
 \caption{As in Fig.~\ref{t14-tc_smarts}, but for the TRAPPIST data. The better contrast 
       (in spite of the larger residual scatter) of the best solution is likely accounted 
       for by the more than ten times larger number of data points than for the SMARTS 
       observations.}
\label{t14-tc_trappist}
\end{figure}

The ($\Delta T_{\rm c}$, $t14$) maps are shown in Figs.~\ref{t14-tc_smarts} 
and \ref{t14-tc_trappist} for the $K_{\rm s}$ and Sloan z' data, respectively. 
As expected, the topology of both maps confirms the rather small (if any) 
deviations from the parameters predicted by the primary transit with the 
assumption of circular orbit. Furthermore, the Sloan z' data are more restrictive 
than the $K_{\rm s}$ data, even though the $S/N$ value is higher for the latter. 
This is because the parameter maps yield information also on the sensitivity of 
the solution on the neighboring parameter values and not only on a specific 
combination of the parameters, that might be better or worse, depending 
on the functional form of the variance on these parameters and noise level. 
The better quality of the Sloan z' data is also visible from the nearly three 
times smaller error of the derived occultation depth.   

%
%
\begin{table}[!h]
  \caption{Observed secondary transit times for WASP-121b}
  \label{occ_times}
  \scalebox{0.9}{
  \begin{tabular}{lrrrc}
  \hline
  Dataset     &  $T_{\rm occ}$ [BJD]  &  $O-C$ [d]   &  $t14$  & Source \\ 
 \hline
SMARTS (K)    &   2457764.6469  & $-0.0080$     & $0.116$     & KOV18 \\
              &   $\pm 0.0023$  & $\pm 0.0023$  & $\pm 0.005$ &       \\
TRAPPIST (z') &   2456762.5594  & $-0.0040$     & $0.128$     & KOV18 \\
              &   $\pm 0.0027$  & $\pm 0.0027$  & $\pm 0.007$ &       \\ 
HST ('white') &   2457703.4588  & $0.0004$      & $-$         & EVA17 \\ 
              &   $\pm 0.0004$  & $\pm 0.0004$  &             &       \\      
Spitzer (3.6) &   2457783.7774  & $-0.0013$     & $-$         & EVA17 \\ 
              &   $\pm 0.0007$  & $\pm 0.0007$  &             &       \\ 
\hline
\end{tabular}}
\begin{flushleft}
{\bf Notes:}\\
\vbox{\small 
EVA17: Evans et al. (2017) -- KOV18: this paper (the source of the 
TRAPPIST data is Delrez et al.~\cite{delrez2016}) -- 
The O-C values are computed in respect of the ephemerides predicted 
from the primary transit as given by Delrez et al.~(\cite{delrez2016}), 
assuming circular orbit. -- See text for the equality of the errors 
on $T_{\rm occ}$ and  $O-C$. -- Evans et al.~(\cite{evans2017}) do not 
supply occultation duration values.   
} 
\end{flushleft}
\end{table}

The currently available secondary transit parameters are summarized in 
Table~\ref{occ_times}. In the case of the KOV18 items the errors have 
been computed in the following way. Once the best-fitting trapezoidal 
was found, we added Gaussian white noise with the observed standard 
deviations of the residuals corresponding to this solution, and 
then the best-fitting trapezoidal to these simulated data was searched 
for. By repeating the process $500$ times we arrived to statistically 
stable estimates of the formal errors. The ingress/egress time was always 
fixed to the observed values given by the primary transit data of 
Delrez et al.~(\cite{delrez2016}), and we did the same also with the 
remaining parameters, depending which parameter was tested for errors 
(e.g., in the case of the occultation center, we fixed the duration and 
the ingress/egress times). 
Although this approach is primarily dictated by keeping the execution 
time within a reasonable limit, our error estimates for the moment of the 
occultation time is in perfect agreement with the one predicted by the 
analytic formula of Deeg \& Tingley~(\cite{deeg2017}). The errors of 
$O-C$ are taken equal to those of $T_{\rm occ}$, because the errors 
of the computed occultation times (C) have been proven to be negligible. 

We see that the available observations suggest a small (or zero) 
eccentricity. Since the more precise estimation requires also the 
knowledge of the transit duration, the lack of this parameter for the 
most accurate HST and Spitzer observations makes us unable to include 
these data in the analysis. Therefore, we use only the occultation 
parameters derived from the SMARTS and TRAPPIST observations. 

Following Winn~(\cite{winn2010}), by omitting the negligible inclination 
effect, we use the following formula to estimate the eccentricity
%
\begin{eqnarray}
\label{eq:6}
e = \Bigg[\Bigg({\pi \over 2}{\Delta T_{\rm c} \over P}\Bigg)^2 + 
         {\Bigg({{r14-1} \over {r14+1}}\Bigg)^2}\Bigg]^{1 \over 2} \hskip 1mm ,
\end{eqnarray}   
where $P$ is the orbital period, $\Delta T_{\rm c}$ is the observed time of 
the occultation center minus the predicted time from the primary transit, 
assuming zero eccentricity; $r14=t14(\rm occ)/t14(\rm tra)$, that is the 
ratio of the secondary and primary transit durations. 
Assuming that the errors are independent on the transit times and durations 
both for the primary and the secondary transits and that these errors are 
also uncorrelated with the error of the period, we can use the above 
equation to estimate the eccentricity and its pure statistical error. 
For the primary transit and period we take the values given in Table~4 
of Delrez et al.~(\cite{delrez2016}). For the secondary transit we use 
the values shown in Table~\ref{occ_times} of this paper. Errors are assumed to be 
Gaussian. Then, Eq.~\ref{eq:6} yields $e=0.0207\pm 0.0153$ if we use 
the SMARTS and $e=0.0314\pm0.0222$, if we use the TRAPPIST data. 
These eccentricity values are also tested by using the primary transit 
center values of Evans et al.~(\cite{evans2018}) for the HST/STIS 
G430Lv2 band (we get very similar results also for the other bands). 
We note that this test is not entirely consistent, since we use the 
transit duration value of Delrez et al.~(\cite{delrez2016}), because, 
Evans et al.~(\cite{evans2018}) do not give this parameter for their 
data. We get for the SMARTS and TRAPPIST data, respectively,  
$e=0.0198\pm 0.0157$ and $e=0.0312\pm0.0224$, i.e., very 
close to those estimated on the basis of the primary transits of 
Delrez et al.~(\cite{delrez2016}). 

In concluding, we note that Delrez et al.~(\cite{delrez2016}) give a 
$3\sigma$ upper limit of $e=0.07$ from the global analysis of the photometric 
and radial velocity data. Our independent analysis is quite consonant with 
theirs.

%
%
\section{Comparison with planet atmosphere models}
\begin{figure}
 \vspace{0pt}
 \centering
 \includegraphics[angle=-90,width=85mm]{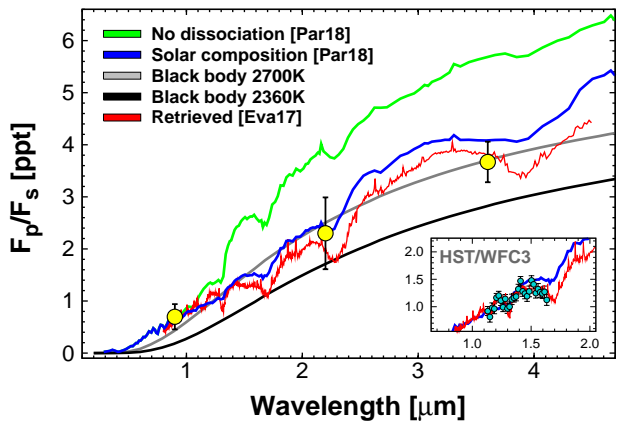}
 \caption{Comparison of the single-band secondary transit depths with 
 the planetary atmosphere models of Parmentier et al.~(\cite{parmentier2018}, 
 [Par18]) and Evans et al.~(\cite{evans2017}, [Eva17]). The error bars show 
 $3\sigma$ statistical uncertainties. We warn that the ``No dissociation'' 
 model is unphysical, and is shown merely to limelight the effect of omitting 
 dissociation in the computation of the spectrum (see text for further details). 
 The black-body lines correspond to different efficiency of the day/night heat 
 transport (black: fully efficient; gray: no heat transport). For completeness, 
 the inset shows the HST observations of Evans et al.~(\cite{evans2017}) with 
 their spectrum retrieval model and the solar composition model of [Par18]. 
 For better visibility, we use $1\sigma$ error bars here.}
\label{models}
\end{figure}

As of the time of this writing, there are the following secondary eclipse 
observations available for WASP-121b. The Sloan z' data at $0.9$~$\mu$m 
by Delrez et al.~(\cite{delrez2016}), the HST data in $1.1-1.6$~$\mu$m 
and the Spitzer data at $3.6$~$\mu$m, both by Evans et al.~(\cite{evans2017}). 
The main panel in Figure~\ref{models} shows the two single-band data 
points together with our occultation depth in the K band at $2.2$~$\mu$m 
(see also Table~\ref{depth_sum} for the actual numerical values used). 
The data are overplotted on the recent planetary atmosphere models of 
Evans et al.~(\cite{evans2017}) and Parmentier et al.~(\cite{parmentier2018}). 
We note that although the ``No dissociation'' model shows very clearly 
that one has to consider element dissociation in modeling HJ atmospheres, 
it is unphysical, and it is shown merely for exhibiting the extreme case 
of neglecting this important physical process. This model was constructed 
by using chemical equilibrium chemistry in the atmospheric structure modul 
of the global circulation model, but $H_2O$ abundance was fixed in computing 
the spectrum. However, the model labelled ``Solar composition'' is consistent 
in this respect, and shows that the currently available data are in an 
overall agreement with it\footnote{By admitting the existence of systematic 
differences for the HST near infrared measurements of Evans et al.~(\cite{evans2017}) 
-- see inset of Fig~\ref{models}.}, without making any special assumption 
or adjustment. Unfortunately, the situation is somewhat more involved, 
as there are several other possibilities yielding spectra rather similar 
to that of the ``Solar composition'' model. For example, one may increase 
the heavy metal content of the ``Solar composition'' model by a factor 
of three, without any essential effect on the emission spectrum -- see 
Parmentier et al.~(\cite{parmentier2018}) for further details.    

%
%
%

%
%
\begin{table}[!h]
  \caption{Secondary transit depths of WASP-121b.}
  \label{depth_sum}
  \scalebox{0.95}{
  \begin{tabular}{lcccc}
  \hline
  Instr./Filter & $\lambda_{\rm c}$ & Depth & Error & Source \\ 
 \hline
 TRAPPIST (z') &  0.9  &  0.697 & 0.081 & Delrez et al.~(\cite{delrez2016}) \\
 2MASS K       &  2.2  &  2.280 & 0.230 & this paper \\
 Spitzer/IRAC  &  3.6  &  3.670 & 0.130 & Evans et al.~(\cite{evans2017}) \\ 
\hline
\end{tabular}}
\begin{flushleft}
{\bf Notes:}\\
\vbox{\small 
Only single-band data are shown. -- The central wavelength $\lambda_{\rm c}$ 
is given in $\mu$m, the depth and its $1\sigma$ statistical error are given 
in ppt (part per thousand). 
} 
\end{flushleft}
\end{table}

The black body lines (gray and black) in Figure~\ref{models} show the 
effect of heat transport from the day-side to the night-side. Assuming zero 
Bond albedo in both cases, the black line displays the case of heat transport 
with maximum efficiency ($A_{\rm B}=0$, $\varepsilon=1.0$ 
-- see Cowan \& Agol~\cite{cowan2011}, Lopez-Morales \& Seager~\cite{lopez2007}). 
It is clear that all available data exclude this possibility and support 
circulation models that are rather inefficient, resulting in a higher day-side 
temperature. For WASP-121b, this temperature seems to be close to $2700$~K, 
corresponding to $\varepsilon=0.57$, assuming $A_{\rm B}=0$. In a comparison 
with the models of Evans et al.~(\cite{evans2017}) -- who use also the above 
planet temperature -- we find that their `retrieved' model slightly 
underestimates our occultation depth by $1.6\sigma$, whereas the mismatch for 
the black body line of $2700$~K is only $0.4\sigma$. 

By scanning the planet temperature, we find that the best-fitting 
black body model to the three single-band data points (weighted equally) 
is reached when $T_{\rm p}=2652$~K. The RMS and the $\chi^2$ value of 
the residuals is $0.194$~ppt and $15.4$, respectively. All points are 
within $1\sigma$, except for the one at $0.9$~$\mu$m, that deviates 
by $3.8\sigma$. For the solar composition model of Parmentier et al.~(\cite{parmentier2018}) 
we get $0.244$~ppt and $10.3$ for the RMS and $\chi^2$, respectively. 
These large values result from the Spitzer data, with a deviation of 
$3.2\sigma$ (the other two points deviate less than $0.5\sigma$). 
Repeating the same comparison for the retrieved model of Evans et al.~(\cite{evans2017}), 
we get, respectively, $0.238$~ppt and $4.7$ for the RMS and $\chi^2$. 
Now all points deviates just barely under $1\sigma$, except for the 
$2.2$~$\mu$m point, that deviates by $1.6\sigma$. 

It is important to note that the status of the outliers might change 
with a different way of handling systematics. As mentioned, over-correcting 
systematics may lead to lower occultation depth (e.g., we got larger depth 
from the $0.9$~$\mu$m data by $\sim0.1$~ppt than the one derived by 
Delrez et al.~\cite{delrez2016}, quite likely, because of the lack of 
polynomial correction in our derivation). 

With the data available today, it seems that the retrieval model of 
Evans et al.~(\cite{evans2017}) is capable to catch most of the features 
of the observed spectrum. The fact that our data deviates by $1.6\sigma$ 
from their model spectrum, indicates that although additional fine tuning 
is needed, the basic characteristics of the data are well-matched. 
On the other hand, the required $VO$ abundance is some thousand times 
of the solar value, which warrants some caution (see 
Evans et al.~\cite{evans2017} and Parmentier et al.~\cite{parmentier2018} 
for further discussion of this issue with the emission spectrum). 
   
Additional complications come from the more extensive data available 
from HST and ground-based transmission spectrum measurements. The 
recent analysis of these data by Evans et al.~\cite{evans2018}) lends 
further support for a high (10--30-times solar) $VO$ abundance and lack 
of $TiO$. Furthermore, these data also pose some challenges in explaining 
the steep rise of the absorption in the near ultraviolet regime. The 
authors invoke sulfanyl ($SH$) as a possible absorber, since the 
standard explanation by Rayleigh scattering fails in the case of 
WASP-121b, due to the high atmospheric temperature implied by Rayleigh 
scattering only. 

Unfortunately, the currently available data on WASP-121b still too 
sparsely populate the more easily measurable part of the emission 
spectrum. In the waveband between $2$~$\mu$m and $4$~$\mu$m (where 
the $CO$ and $H_2O$ emissions are the most pronouncing) additional 
data would be of great help. High S/N measurements carried out by 
instruments like CRIRES at VLT would be clearly capable to map this 
crucial region. In addition to the determination of the abundances 
of the molecules above, this would perhaps constrain also the abundances 
derived from the shorter wavelength part of the spectrum, where 
high S/N data gathering is more difficult.

%
%
\section{Conclusions}
We presented the first secondary transit measurements of an extrasolar 
planet in the near infrared by using a 1-m class telescope. With the 
ANDICAM imager attached to the $1.3$~m telescope of the SMARTS Consortium, 
we detected an occultation depth of $(0.228\pm 0.023)$\% in the 2MASS 
K band from observations made in three nights on the very hot Jupiter 
WASP-121b. We compared this value with theoretical planetary spectra of 
Parmentier et al.~(\cite{parmentier2018}) and Evans et al.~(\cite{evans2017})  
and found that it fits perfectly the former model, using solar composition, 
atmospheric circulation and molecular dissociation. However, when considering 
all available secondary transit data (Sloan z', HST and Spitzer data -- see 
Delrez et al.~\cite{delrez2016} and Evans et al.~\cite{evans2017}), 
it seems that the $VO$-enhanced model of Evans et al.~(\cite{evans2017}) is 
preferred over the solar composition model, albeit with a less favorable 
match to our data. Although the $2700$~K black body line yields also an 
acceptable overall fit to the available data, the more detailed HST 
spectrum is not reproduced well. Additional data in the $(2$ -- $4)$~$\mu$m 
regime would be very useful to verify model predictions on $CO$ and $H_2O$ 
emissions and build a more coherent planet atmosphere model. 

%
%
\begin{figure}
 \vspace{0pt}
 \centering
 \includegraphics[angle=-90,width=85mm]{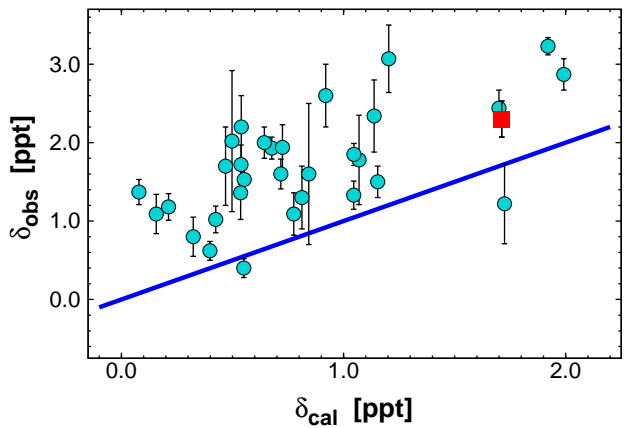}
 \caption{Observed vs calculated secondary transit depths for the $32$ extrasolar 
       planets known today with emission measurements at $\sim 2.2$~$\mu$m. 
       Nearly all observations lie above the equality line, corresponding 
       to the calculated/expected black body value, assuming effective heat 
       transport from the day side to the night side. WASP-121b is shown 
       as a red square. Error bars show $1\sigma$ statistical errors.}
\label{o_vs_c_depth}
\end{figure}

In agreement with other studies (e.g., Adams \& Laughlin~\cite{adams2018}, 
and references therein), our data support the lack of efficient day-to-night 
side heat transport (see Fig.~\ref{models}). This conclusion is further 
strengthened if we compare the predicted and observed occultation depths by 
using all available data today. Based on the list of Alonso~(\cite{alonso2018}), 
we collected the secondary transit depths measured in the 2MASS K band for 
$32$ Hot Jupiters (see Croll et al.~\cite{croll2015}, Cruz et al.~\cite{cruz2015}, 
Zhou et al.~\cite{zhou2015}, Martioli et al.~\cite{martioli2018} and this 
paper). The observed depths as a function of the expected value (assuming 
zero Bond albedo and fully efficient heat transport) are shown in  
Fig.~\ref{o_vs_c_depth}. The figure clearly shows a nearly uniform offset, 
with no apparent dependence on the expected depth. The effect is 
exacerbated if we consider more realistic albedos as suggested by recent 
analyses of full orbit phase curves -- see Adams \& Laughlin~(\cite{adams2018}).   

We arrive to a similar conclusion if we examine the difference between 
the observed and calculated occultation depths as a function, e.g., of 
the temperature at the substellar point. Therefore, -- admitting 
the need for a more complete characterization of the heat distribution 
by directly measuring the night- and day-side fluxes 
(i.e., Komacek \& Showman~\cite{komacek2016}) -- from the $2.2$~$\mu$m 
measurements alone, there does not seem to exist a correlation between 
the heat redistribution efficiency and planet temperature 
(i.e., Cowan \& Agol~\cite{cowan2011}, 
{Komacek \& Showman~\cite{komacek2016}). Supporting our result, 
it is interesting to note that a similar study by 
Baskin et al.~(\cite{baskin2013}), based on Spitzer $3.6$~$\mu$m and 
$4.5$~$\mu$m data has led to the same conclusion. 

Although our observations were made in a single waveband, they yield a 
reasonably solid piece of information both on the orbital and on 
the atmospheric characterization of the WASP-121 system. Together with 
future emission data in the $(2$ -- $4)$~$\mu$m band they will allow to 
prove or deny the existence of the $CO$, $H_2O$ emission feature 
on the day side predicted by the models in this waveband.

%
%
\begin{acknowledgements}
We thank to Laetitia Delrez for sending us the secondary transit observations 
presented in the discovery paper on WASP-121. We are grateful to Vivien 
Parmentier for making the relevant planet atmosphere models accessible to us 
and helping in the comprehension of the models. The professional help given 
by the SMARTS staff at the Yale University during the data acquisition period 
is much appreciated. We also thank the referee for the critical notes on our 
early interpretation of the planet atmosphere models. The observations have 
been supported by the Hungarian Scientific Research Fund (OTKA, grant K-81373). 
TK acknowledges the support of Bolyai Research Fellowship. Additional grants 
(PD~121223 and K~129249) from the National Research, Development and Innovation 
Office are also acknowledged. 
\end{acknowledgements}

\bibliographystyle{aa} 

\end{document}